\documentclass{aastex631}
\usepackage{graphicx}

\shorttitle{Blocking off the BH growth at high-$z$}
\shortauthors{Kawanaka \& Kohri}
\graphicspath{{./}{figures/}}
\begin{document}

\title{Effects of Heat Conduction on Blocking off the Super-Eddington Growth of Black Holes at High Redshift}

\author{Norita Kawanaka\thanks{Present address: Department of Physics, Graduate School of Science Tokyo Metropolitan University 1-1,
Minami-Osawa, Hachioji-shi, Tokyo 192-0397}}
\affiliation{Center for Gravitational Physics and Quantum Information, Yukawa Institute for Theoretical Physics, Kyoto University, \\
Kitashirakawa Oiwake-cho, Sakyo-ku, Kyoto, 606-8502, Japan}

\author{Kazunori Kohri}
\affiliation{Theory Center, Institute of Particle and Nuclear Studies, KEK, Tsukuba 305-0801, Japan}
\affiliation{QUP, KEK, Tsukuba 305-0801, Japan}
\affiliation{Department of Particles and Nuclear Physics, the Graduate University for Advanced Studies (Sokendai), Tsukuba 305-0801, Japan}
\affiliation{Kavli IPMU (WPI), UTIAS, The University of Tokyo, Kashiwa, Chiba 277-8583, Japan}



\begin{abstract}
We investigate the effect of conductive heating of the gas surrounding a geometrically thick accretion disk on the growth of a black hole at high redshift.  If a black hole is accreting the surrounding gas at a super-Eddington rate, the X-ray radiation from the vicinity of the black hole would be highly anisotropic due to the self-shielding of a geometrically thick accretion disk, and then the radiative feedback on the surrounding medium would be suppressed in the equatorial region, within which super-Eddington accretion can continue.  However, if this region is sufficiently heated via thermal conduction from the adjacent region that is not shielded and heated by the X-ray irradiation, the surrounding gas becomes isotropically hot and the Bondi accretion rate would be suppressed and become sub-Eddington.  We evaluate the condition under which such isotropic heating is realized, and derive new criteria required for super-Eddington accretion.
\end{abstract}



\section{Introduction}
The formation and growth of massive black holes at early epochs have been long-standing mysteries.  The most distant quasar detected to date, J0313-1806, lies at a redshift $z=7.642$, corresponding to $\lesssim 6.7\times 10^8~{\rm yr}$ after the big bang, and is believed to have a black hole with a mass of $(1.6\pm+0.4)\times 10^9M_{\odot}$ \citep{2021ApJ...907L...1W}.  This observational fact poses a challenge to the theoretical models of black hole growth (see \citealp{2020ARA&A..58...27I} for a recent review).  In fact, if the mass accretion onto a black hole is limited below the Eddington rate, the growth of a light `seed' black hole ($M_{\rm BH}\lesssim 100M_{\odot}$) that is plausibly formed at $z\gtrsim 20$ to such a massive black hole requires the uninterrupted accretion lasting at least for $\gtrsim 6\times 10^8~{\rm yr}$ \citep{madau+14} and the relatively small mass-to-light conversion efficiency ($\epsilon\lesssim 0.1$; \cite{tanakahaiman09}).

To explain the existence of such supermassive black holes (SMBH) in the high redshift universe, two ideas are frequently discussed in the literature: one is to produce a massive seed black hole with mass of $\gtrsim 10^5M_{\odot}$ from the direct collapse of a supermassive star by general relativistic instability \citep{loebrasio94, 2001ApJ...546..635O, ohhaiman02, brommloeb03, lodatonatarajan06, begelman+06, spaanssilk06, inayoshiomukai12, regan+14, inayoshi+14, ferrara+14, becerra+15, latif+15, 2018MNRAS.475.4104C, 2019Natur.566...85W, 2021MNRAS.506..613S} or by runaway stellar collisions in a stellar cluster \citep{sanders70, portegieszwart+04, freitag+06, omukai+08, devecchivolonteri09, katz+15, yajimakhochfar15, 2017MNRAS.472.1677S, 2020ApJ...892...36T, 2020MNRAS.494.2851C}.  With such a massive seed, assuming the gas accretion at the Eddington rate, the timescale required to form $\gtrsim 10^9M_{\odot}$ supermassive black holes would be reduced.

The other scenario is the super-Eddington accretion onto a light seed black hole  \citep{volonterirees05, wyitheloeb12, madau+14, alexandernatarajan14, volonteri+15, lupi+15}.  In general the mass accretion exceeding the Eddington rate is accompanied with strong radiation force outward, which may prevent the infall of surrounding gas.  However, this difficulty may be avoided by two different ways.  The first one is to consider the slim disk model \citep{abramowicz+88}.  Slim disks are the solutions of accretion flows with mass accretion rate around and above the Eddington rate ($\dot{M}\gtrsim L_{\rm Edd}/c^2$), in which photons are almost completely trapped and thus the advective cooling dominates over the radiative cooling.  The evidence of such accretion flows existing around black holes has been accumulating.  For example, the observed properties of some ultraluminous X-ray sources (ULXs; $L_X \gtrsim 10^{39}~{\rm erg}~{\rm s}^{-1}$) can be explained by the model of a slim disk around a stellar-mass black hole \citep{2001ApJ...552L.109K, 2001ApJ...549L..77W, gladstone+09, middleton+13}.  As for active galactic nuclei (AGNs), some of narrow-line Seyfert 1 galaxies (NLS1) are believed to be energized by super-Eddington accretion onto a relatively less massive black hole \citep{mineshige+00, kawaguchi03}.  Recent radiation (magneto-)hydrodynamical simulations have shown the feasibility of the super-Eddington mass accretion onto a black hole \citep{2009PASJ...61L...7O, ohsugamineshige11, 2014ApJ...796..106J, 2014MNRAS.441.3177M, 2015MNRAS.447...49S, 2016ApJ...826...23T, 2019ApJ...880...67J}.  However, one should take into account the strong feedback from an accretion flow onto the surrounding medium due to X-ray radiation and mass outflow, which are inevitable in the super-Eddington phase \citep{ciottiostriker01, wang+06, milosavlijevic+09, alvarez+09}.  \citet{2016MNRAS.459.3738I} discussed the spherically symmetric super-Eddington accretion onto a seed black hole instead of disk-like accretion.  They concluded that the steady hyper-Eddington accretion is realized because of photon trapping, which makes any radiative feedback ineffective, and that a black hole can grow up to a maximum $\sim 10^5~M_{\odot}$ independent of its initial mass.

In a realistic case, however, the ambient gas around a BH has nonzero angular momentum, which makes the mass accretion onto a BH rather anisotropic and disk-like.  The super-Eddington growth of a seed black hole with disk-like accretion has been investigated in some studies \citep{2017MNRAS.469...62S, 2018MNRAS.476..673T}.  In such a case, the radiation flux is collimated towards the bipolar directions, where the ambient gas is efficiently heated and the Bondi accretion rate would be suppressed because it is proportional to $c_s^{-3}$ where $c_s$ is the speed of sound in the ambient gas.  However, when the mass accretion rate is close to or exceeds the Eddington value, the disk becomes geometrically thick, and therefore the ambient medium around the equatorial plane would be able to avoid the strong irradiation from the inner part of the accretion flow because of the self-shielding effect \citep{watarai+05}.  This means that even without the assumption of spherical symmetry we can expect the mass accretion that does not suffer from any radiative feedback.  In addition to this effect, \citet{2019MNRAS.488.2689T} investigated the effect of the radiation spectra from an accretion disk by performing two-dimensional multifrequency radiation hydrodynamical simulations, and obtain the criterion required for super-Eddington accretion.  It has also been pointed out that the powerful outflow driven by radiation pressure would affect the growth rate \citep{2020MNRAS.497..302T}.  In the numerical simulations performed in these studies, only the radial component of the radiation flux is calculated.  Actually, the ambient gas is fully ionized and optically thin with respect to electron scattering, and so the non-radial propagation of photons can be neglected.  However, there should be non-radial heat propagation due to conduction, which is not taken into account in the simulations above.  If the conduction is efficient enough that the ambient gas at the Bondi radius is heated up isotropically, the mass accretion from any direction would be suppressed.

In this study, we reexamine the criterion required for super-Eddington accretion onto a seed BH taking into account the disk-like accretion as well as non-radial heating of the ambient gas due to conduction.  In Section 2 we briefly review the growth of a seed BH due to the Bondi accretion in the high-redshift universe.  We describe our model of the disk-like accretion onto a seed BH that take into account the thermal conduction in Section 3, and show the results in Section 4.  In Section 5 we discuss the implication of our results as well as the difference from previous studies, and we summarize our work in Section 6.

\section{Super-Eddington accretion onto a seed BH in a pregalactic halo}
\citet{volonterirees05} argued the evolution of the seed of a supermassive black hole (SMBH) with the mass of $M_{\rm BH}$ hosted by metal-free pregalactic halos with virial temperature $T_{\rm vir}>10^4~{\rm K}$.  In such a halo, due to the lack of ${\rm H}_2$ molecules, gas can cool via hydrogen atomic lines to $\sim 8000~{\rm K}$.  This halo generally has angular momentum, $J$, which is related to what is called the spin parameter $\lambda=J|E|^{1/2}/GM_h^{5/2}$, where $E$ and $M_h$ are the total energy and mass of the halo, respectively.  Following \citet{ohhaiman02}, they assume that a fraction of the gas, $f_d$, forms an isothermal exponential disk whose scale radius is given by $R_d \sim 2^{-1/2}\lambda R_{\rm vir}$.  In this disk, the number density of hydrogen at radius $r$ and at vertical height $z$ can be written as
\begin{eqnarray}
n(r,z)=n_0\exp \left( -\frac{2r}{R_d} \right) {\rm sech} ^2\left( \frac{z}{\sqrt{2}z_0} \right),
\end{eqnarray}
where $n_0$ is the central density, $z_0$ is the vertical scale height at radius $r$,
\begin{eqnarray}
z_0=\frac{c_s}{\left( 4\pi G \mu m_{\rm H} n_0 e^{-2r/R_d} \right)^{1/2}},
\end{eqnarray}
$c_s$ is the sound speed of the gas, $\mu=0.6$ is the mean molecular weight.

Since the mass of the disk can be written as $M_d=f_d(\Omega_b/\Omega_M)M_h$, the central number density of the gas can be written as
\begin{eqnarray}
n_0\simeq 6\times 10^4 f_{d,0.5}^2\lambda_{0.05}^{-4}T_{{\rm gas},8000}^{-1}R_{{\rm vir},6}^{-4}M_{h,9}^2~{\rm cm}^{-3},
\end{eqnarray}
where $M_{h,9}=M_h/10^9M_{\odot}$, $T_{{\rm gas},8000}$ is the gas temperature in units of $8000~{\rm K}$, $R_{{\rm vir},6}$ is the virial radius in units of $6~{\rm kpc}$, $f_{d,0.5}=f_d/0.5$, and $\lambda_{0.05}=\lambda/0.05$.

Let us consider how a seed BH grows in this halo.  In a spherical geometry, matter would be accreted into a BH at the rate derived by \citet{bondi52},
\begin{eqnarray}
\dot{M}_{\rm Bondi}=\frac{4\pi G^2 M_{\rm BH}^2m_p n_0}{c_s^3},
\end{eqnarray}
and the radius inside which gas would be accreted by a central black hole (Bondi radius) is described as
\begin{eqnarray}
r_{\rm Bondi}=\frac{GM_{\rm BH}}{c_s^2}\simeq 2.0\times 10^{17}~{\rm cm}~M_{{\rm BH},3} T_{{\rm gas},8000}^{-1},
\end{eqnarray}
where $M_{{\rm BH},3}=M_{\rm BH}/10^3M_{\odot}$, $r_g=2GM_{\rm BH}/c^2$ is the Schwarzshild radius of a black hole, $c_s$ is the speed of sound, and $T_{\rm gas}$ is the temperature of the surrounding gas ($T_{{\rm gas},8000}=T_{\rm gas}/8.0\times 10^3~{\rm K}$).  On the other hand, the Eddington accretion rate can be described as $\dot{M}_{\rm Edd}=4\pi GM_{\rm BH}m_p/(\eta \sigma_T c)$, where $\eta=1/16$ is the efficiency of BH accretion in the pseudo-Newtonian potential \citep{paczynskiwiita80}, and $\sigma_T$ is the Thomson scattering cross section.  Then comparing the Bondi accretion rate to the Eddington rate, we can see that the mass accretion would naturally be super-Eddington:
\begin{eqnarray}
\frac{\dot{M}_{\rm Bondi}}{\dot{M}_{\rm Edd}}=3.1~M_{{\rm BH},3}n_{0,4}T_{{\rm gas},8000}^{-3/2}.
\end{eqnarray} 

\section{Model}
\subsection{Radiative Feedback}
As long as the condition discussed in \citet{volonterirees05} is satisfied, we can expect super-Eddington accretion onto a seed BH through a tiny accretion disk.  However, they do not consider the radiative feedback from the accretion disk to the surround gas.  Actually, a huge amount of ionizing photons would come out from the inner part of the accretion disk and heat the ambient gas.  As a result, it is expected that the pressure gradient force of the heated gas would exceed the gravitational force, which would suppress the accretion from larger radii \citep{ciottiostriker01, alvarez+09, milosavlijevic+09}.  \citet{2016MNRAS.459.3738I} discussed the case in which a seed BH is embedded in a sufficiently dense gas, and showed that the ionized region would collapse due to intense inflows of neutral gas, which enables an isothermal Bondi accretion with hyper accretion rate, $\sim 500\dot{M}_{\rm Edd}$.  They also derived the criterion for this hyper-Eddington accretion as
\begin{eqnarray}
n_0 \gtrsim 10^4~{\rm cm}^{-3}~M_{{\rm BH},5}T_{{\infty},4}^{3/2},
\end{eqnarray}
when the system is spherically symmetric (see also \citealp{2016MNRAS.461.4496S}).  In a realistic case, however, the radiation flux from the vicinity of a BH is anisotropic due to the self-shielding by a geometrically thick accretion disk.  \citet{2017MNRAS.469...62S} and \citet{2018MNRAS.476..673T} independently investigated the gas accretion in such cases by two-dimensional radiation hydrodynamical simulations, and showed that rapid accretion through the equatorial plane is still possible, while the ionized gas can expand towards the rotation axis producing hot outflows with $T\sim 10^5~{\rm K}$.

In the ionized region, radiative heating due to Compton scattering would be also effective.  \citet{wang+06} discussed the effects of Compton heating of the ambient medium by hard X-ray photons emitted from the inner part of the accretion disk, and showed that the radius of Compton-heated region would be larger than the Bondi radius, which would reduce the Bondi accretion rate far below the Eddington rate.  Especially, the temperature of an accretion disk corona producing hard X-ray photons would be several tens of keV in the super-Eddington regime \citep{2021PASJ...73..630K}, which can heat the ambient medium up to $\sim 10^8~{\rm K}$.

In the studies mentioned above, they do not consider the possibility that the gas in the shielded region can be heated somehow.  First, one can consider the heating by the non-radial radiation flux that may be generated by electron scatterings in the irradiated region.  However, since the irradiated gas is optically thin with respect to electron scattering \citep{2018MNRAS.476..673T}, we can neglect this process.  Second, since the irradiated hot region and the shielded cool region is adjacent, one can expect the conductive heat transfer from the former to the latter.  In fact, this heating process is not negligible in some situations.  In the next subsection we discuss the condition under which the conductive heat transfer becomes significant and derive new criteria for super-Eddington accretion taking into account this effect.

\subsection{Thermal Conduction in the Shielded Region}
\begin{figure}[tbp]
\begin{center}
\includegraphics[width=10cm, clip]{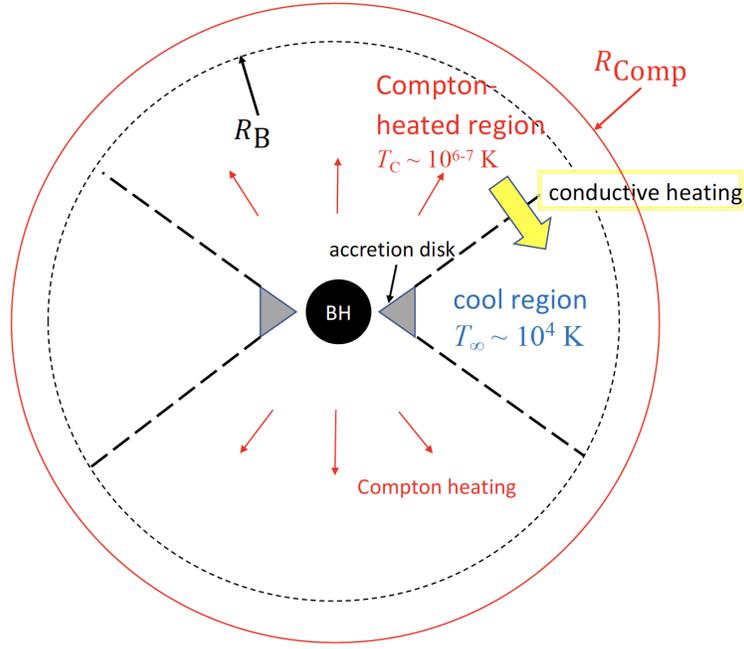}
\caption{Schematic picture of the thermal structure around a Bondi-accreting seed black hole.  $R_{\rm B}$ and $R_{\rm Comp}$ are the Bondi radius and Compton radius (see the text), respectively.}
\label{schematic}
\end{center}
\end{figure}

In this subsection, we consider the effect of thermal conduction from the heated region (i.e., non-shielded region) to the cool region (i.e., shielded region).  Fig. \ref{schematic} depicts the thermal structure of the ambient gas around an accreting seed BH.  If the shielded region is heated up enough due to the thermal conduction from the heated region, the ambient medium is isotropically heated, which makes the Bondi accretion rate suppressed and therefore a BH cannot grow so fast.  Here we discuss the conditions that the conduction heating becomes efficient.

First, let us consider the Compton heating of the ambient gas that is fully ionized by the irradiation from the inner disk following \citet{2014MNRAS.445.2325P}. The Compton heating rate per volume can be described as
\begin{eqnarray}
\Gamma_{\rm C}=\frac{\sigma_T n_e}{m_e c^2}\int_{{\nu}_{\rm min}}^{{\nu}_{\rm max}} d\nu F_{\nu}(h\nu -4k_{\rm B}T),
\end{eqnarray}
where $n_e$, $T$, $\nu_{\rm min}$ ($\nu_{\rm max}$) and $F_{\nu}$ are the ambient electron number density, ambient temperature, the minimum (maximum) frequency of the irradiation, and the energy flux of the irradiation per frequency, respectively.  Assuming that the spectrum of the irradiation energy flux can be approximated by a power-law with an index of $\alpha$, the Compton heating rate can be rewritten as
\begin{eqnarray}
\Gamma_{\rm C}=F\frac{\sigma_T n_e}{m_e c^2}\left(\frac{h\nu_{\rm max}}{\Lambda_{\rm C}}-4k_{\rm B}T\right), \label{comptonheating}
\end{eqnarray}
where $F$ is the frequency-integrated energy flux of the irradiation, and $\Lambda_{\rm C}$ is defined by the spectral shape of the irradiating photons as
\begin{eqnarray}
\Lambda_{\rm C}=\left( \frac{2-\alpha}{\alpha-1} \right)\frac{\nu_{\rm max}^{2-\alpha}-\nu_{\rm min}^{2-\alpha}(\nu_{\rm max}/\nu_{\rm min})}{\nu_{\rm max}^{2-\alpha}-\nu_{\rm min}^{2-\alpha}}.
\end{eqnarray}
Note that this expression is valid only when $\alpha>1$.  The irradiation field and the gas are in equilibrium when $\Gamma_{\rm C}=0$, which means that the gas is heated up to the temperature of
\begin{eqnarray}
T_{\rm C}=\frac{h\nu_{\rm max}}{4k_{\rm B}\Lambda_C},
\end{eqnarray}
which we call as the Compton temperature.

Let estimate the maximum radius within which the gas around a black hole can be heated up to the Compton temperature (hereafter we call this radius as the Compton radius).  There are two different definitions for the Compton radius.  One is defined using the critical ionization parameter $\Xi_{\rm c}=1.1\times 10^3 T_{\rm C,6}^{-3/2}$ \citep{krolik+81} as
\begin{eqnarray}
R_{\rm Comp}^{\Xi}&=&\left(\frac{L}{4\pi c \Xi_{\rm c}nk_B T}\right)^{1/2}\nonumber \\
&\simeq&2.7\times 10^{18}~{\rm cm}~M_{\rm BH,5}^{1/2}\lambda_{\rm Edd}^{1/2}T_{\rm C,7}^{3/4}n_{\infty,5}^{-1/2}T_{\infty,4}^{-1/2},
\end{eqnarray}
beyond which Compton heating is no longer efficient.  Here, $\lambda_{\rm Edd}=L/L_{\rm Edd}$ is the Eddington ratio of the irradiation luminosity.  

The other definition comes from the requirement that the Compton heating timescale, $t_{\rm C}=6\pi m_e c^2 \left(R_{\rm Comp}^{\Xi}\right)^2/\sigma_T L$, should be shorter than the accretion timescale, $t_{\rm infall}=R^{3/2}/\left(2GM_{\rm BH}\right)^{1/2}$:
\begin{eqnarray}
R_{\rm Comp}^{\rm infall}&=&\frac{1}{2GM_{\rm BH}}\left( \frac{\sigma_T L}{6\pi m_e c^2}\right)^2 \nonumber \\
&\simeq& 1.2\times 10^{20}~{\rm cm}~M_{\rm BH,5}\lambda_{\rm Edd}^2.
\end{eqnarray}
Therefore, we should define the Compton radius as $R_{\rm Comp}={\rm min}(R_{\rm Comp}^{\Xi},R_{\rm Comp}^{\rm infall})$.

For the suppression of mass accretion rate, it is required that the Compton radius is larger than the Bondi radius because otherwise the gas in between the former and the latter would accumulate and crush the hot bubble inside \citep{2016MNRAS.459.3738I, 2017MNRAS.469...62S}, which makes hyperaccretion onto a black hole possible.  One can rewrite the condition $R_{\rm Comp}^{\Xi}>R_{\rm Bondi}$ as
\begin{eqnarray}
\eta_{-2}^{1/2}T_{\rm C}^{3/4}T_{\infty,4}^{-1/4}\gtrsim 0.322, \label{rcomprb1}
\end{eqnarray}
and the condition $R_{\rm Comp}^{\rm infall}>R_{\rm Bondi}$ as
\begin{eqnarray}
\eta_{-2}^2 T_{\infty,4}^{-2}n_{\infty,5}^2M_{\rm BH,5}^2\gtrsim 6.3\times 10^{-7}. \label{rcomprb2}
\end{eqnarray}
Here $\eta=10^{-2}\eta_{-2}=L/(\dot{M}_{\rm Bondi}c^2)$ is the radiative efficiency of the accretion flow.

Next, let us derive the condition with which the equatorial region that is shielded by a geometrically-thick accretion disk is efficiently heated by the thermal conduction from the irradiated region.  Assuming that the aspect ratio of the accretion disk is $H/r\sim 1$ \citep{2005ApJ...629..341K}, the thermal conduction flux from the irradiated region to the shielded region at a distance $r$ from a black hole can be described as
\begin{eqnarray}
F_{\rm cond}&=&\kappa_{\rm sp}\frac{\partial T}{\partial z} \simeq \kappa_{\rm sp}\frac{T_{\rm C}}{r}, \label{conductionflux}
\end{eqnarray}
where $\kappa_{\rm sp}\simeq 10^{-6}T_{\rm C}^{5/2}~{\rm ergs}~{\rm cm}^{-1}~{\rm s}^{-1}~{\rm K}^{-1}$ is the Spitzer's heat conductivity \citep{spitzer65}.  Therefore, the heating timescale of the shielded region at a certain radius $r$ can be expressed as
\begin{eqnarray}
t_{\rm cond}(r)&\simeq & \frac{n_{\infty}k_{\rm B}T_{\rm C}\cdot r}{\kappa_{\rm sp}T_{\rm C}/r} \nonumber \\
&\simeq& 2.4\times 10^9~{\rm s}~n_{\infty,5}T_{\rm C,7}^{-5/2}r_{16}^2,
\end{eqnarray}
while the accretion timescale is expressed as
\begin{eqnarray}
t_{\rm acc}&=&\sqrt{\frac{r^3}{GM_{\rm BH}}} \nonumber \\
&\simeq&8.66\times 10^9~{\rm s}~M_{\rm BH,2}^{-1/2}r_{16}^{3/2}.
\end{eqnarray}
From the condition $t_{\rm acc}(R_{\rm Bondi}) \lesssim t_{\rm cond}(R_{\rm Bondi})$, we can evaluate the condition that the shielded region is efficiently heated by heat conduction as
\begin{eqnarray}
M_{\rm BH,5}n_{\infty,4}T_{\infty,4}^{-1/2}T_{\rm C,7}^{-5/2}\gtrsim 2.57\times 10^{-2}. \label{tcompr}
\end{eqnarray}
When the conditions Eqs. (\ref{rcomprb1}) (or \ref{rcomprb2}) and (\ref{tcompr}) are fulfilled at the same time, the mass accretion rate from any solid angle would be the Bondi value with temperature of $T_{\rm C}$,
\begin{eqnarray}
\frac{\dot{M}_{\rm Bondi,C}}{\dot{M}_{\rm Edd}}\simeq 7.0\times 10^{-3}M_{\rm BH,3}n_{0,4}T_{\rm C,7}^{-3/2},
\end{eqnarray}
which means that the growth of the black hole would be suppressed below the Eddington rate.

\section{Results}
\begin{figure}[tbp]
\begin{center}
\includegraphics[width=10cm]{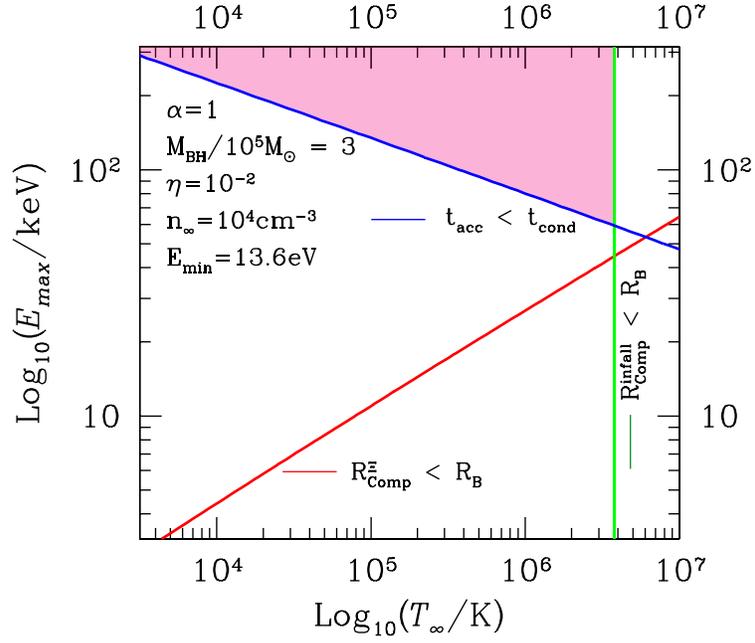}
\caption{The parameter region in which the mass accretion onto a seed BH would be suppressed on the plane of ambient temperature $T_{\infty}$ and the maximum photon energy of the irradiation flux $E_{\rm max}$ (shaded region).  The spectral index of the irradiation flux and the black hole mass are set as $\alpha=1$ and $M_{\rm BH}=3\times 10^5M_{\odot}$, respectively.}
\label{alpha1mbh3e5}
\end{center}
\end{figure}

\begin{figure}[tbp]
\begin{center}
\includegraphics[width=10cm]{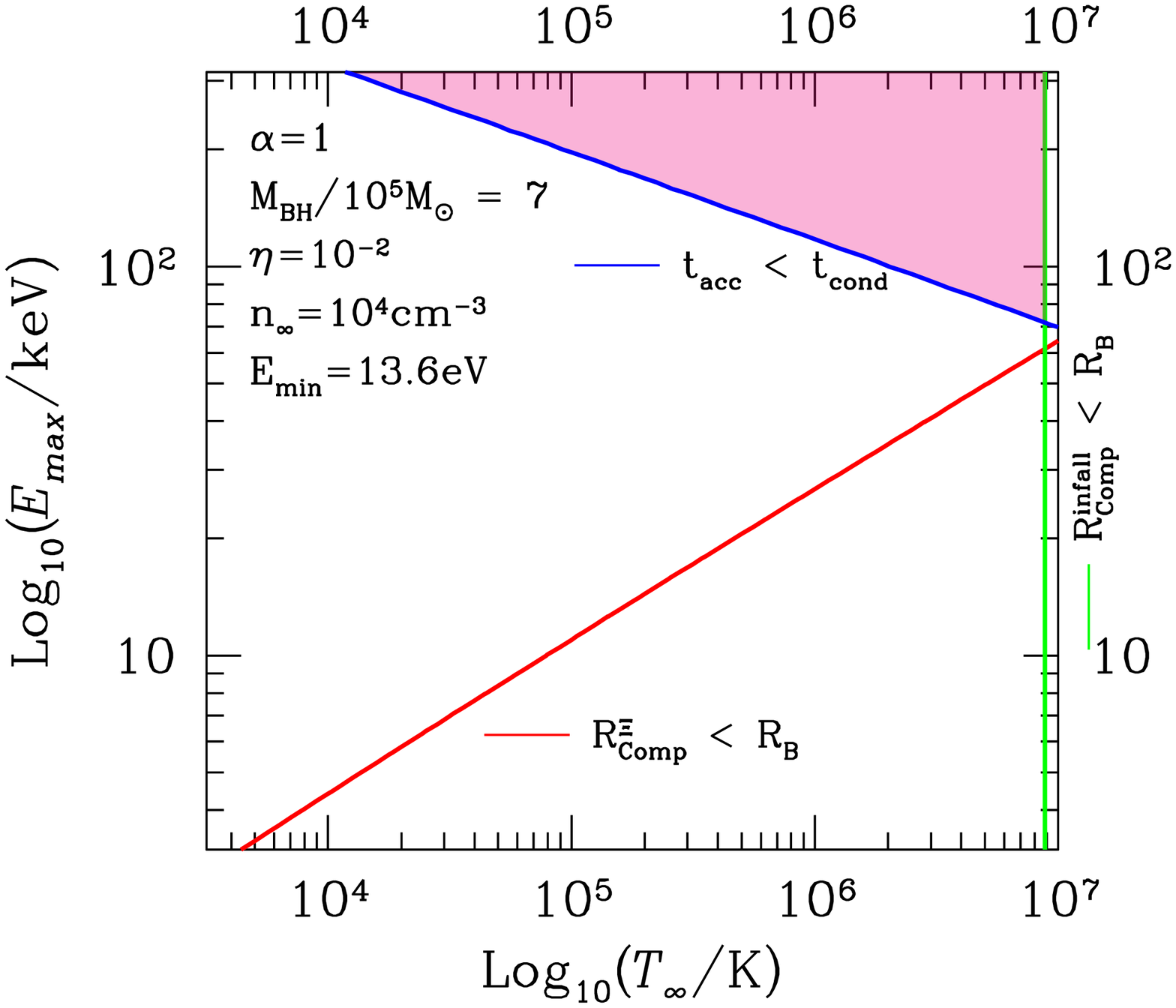}
\caption{Same as Fig. \ref{alpha1mbh3e5}, but with $\alpha=1$ and $M_{\rm BH}=7\times 10^5M_{\odot}$.}
\label{alpha1mbh7e5}
\end{center}
\end{figure}

\begin{figure}[tbp]
\begin{center}
\includegraphics[width=10cm]{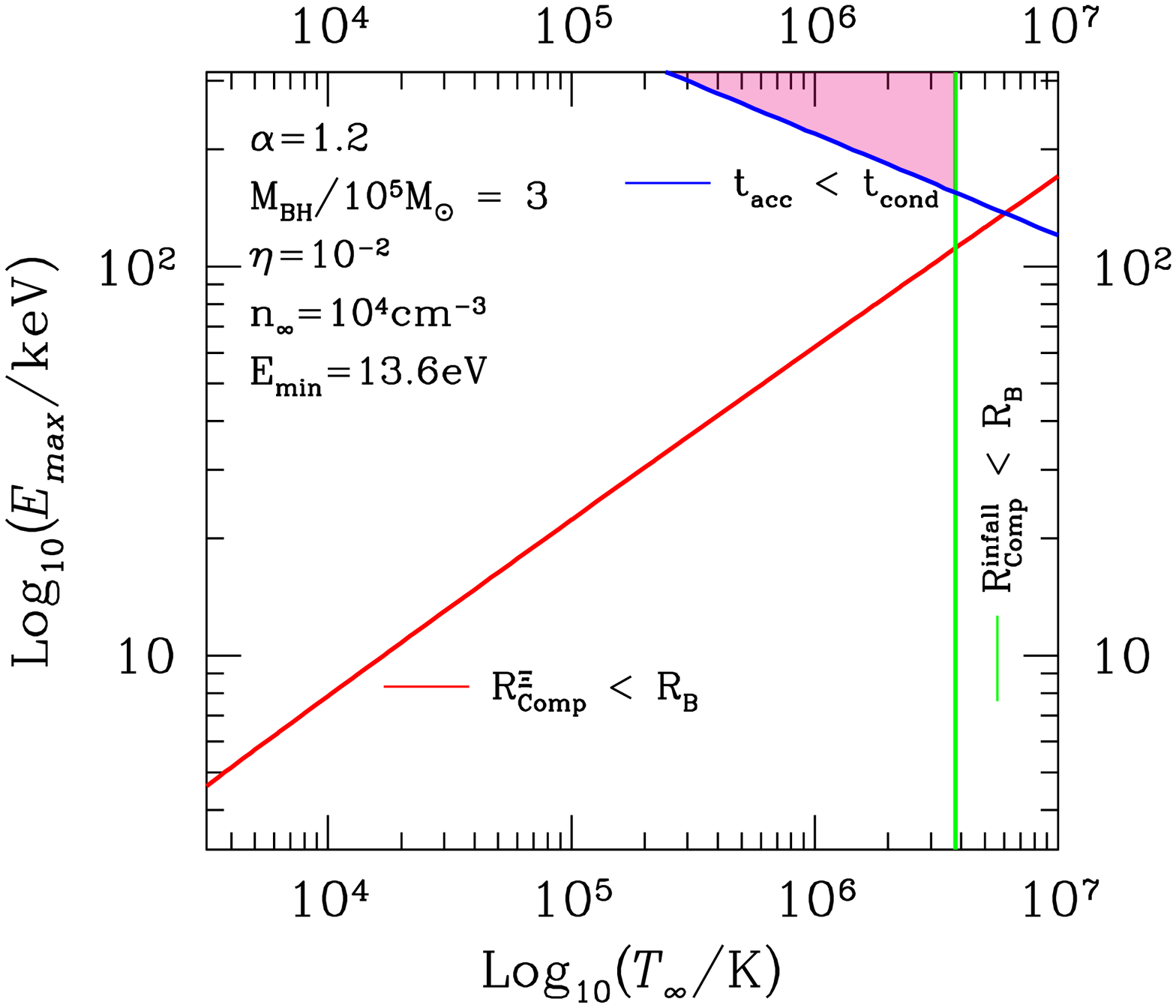}
\caption{Same as Fig. \ref{alpha1mbh3e5}, but with $\alpha=1.2$ and $M_{\rm BH}=3\times 10^5M_{\odot}$.}
\label{alpha1.2mbh3e5}
\end{center}
\end{figure}

\begin{figure}[tbp]
\begin{center}
\includegraphics[width=10cm]{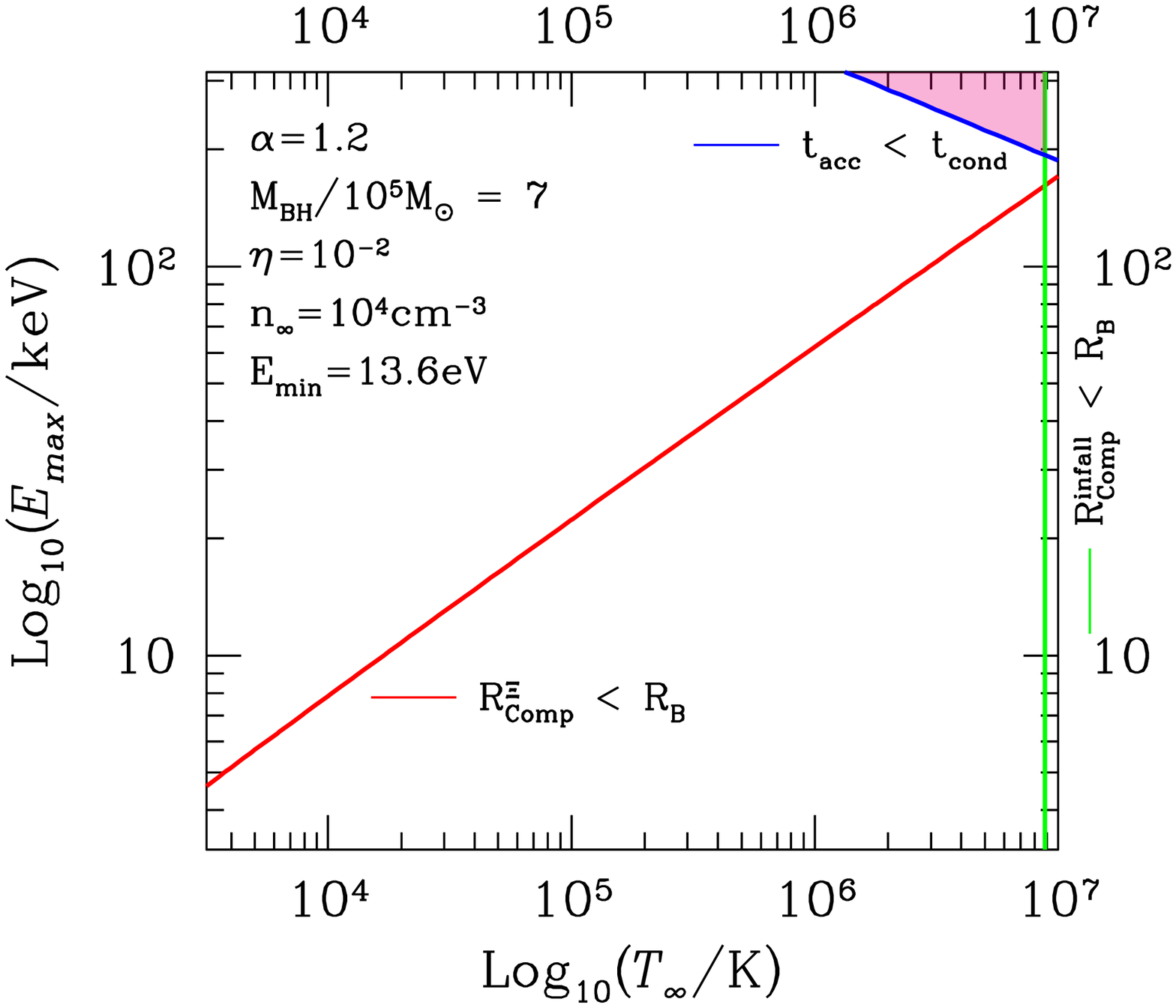}
\caption{Same as Fig. \ref{alpha1mbh3e5}, but with $\alpha=1.2$ and $M_{\rm BH}=7\times 10^5M_{\odot}$.}
\label{alpha1.2mbh7e5}
\end{center}
\end{figure}

In this section we show the conditions under which the super-Eddington growth of a seed BH is suppressed due to heat conduction.  Fig. \ref{alpha1mbh3e5} depicts the parameter region in which the super-Eddington BH growth is suppressed on the plane of the temperature of the ambient gas ($T_{\infty}$) and the maximum photon energy of the irradiation flux ($E_{\rm max}=h\nu_{\rm max}$) by a shaded region with $\alpha=1$, $M_{\rm BH}=3\times 10^5M_{\odot}$, $n_{\infty}=10^4~{\rm cm}^{-3}$, and $E_{\rm min}=13.6~{\rm eV}$.  Fig. \ref{alpha1mbh7e5}, Fig. \ref{alpha1.2mbh3e5}, and Fig. \ref{alpha1.2mbh7e5} are the same as Fig.1, but with $(\alpha, M_{\rm BH})=(1, 7\times 10^5M_{\odot})$, $(1.2, 3\times 10^5M_{\odot})$, and $(1.2, 7\times 10^5M_{\odot})$, respectively.  One can see that the super-Eddington accretion is suppressed when $E_{\rm max}$ is high enough, and that the condition for $E_{\rm max}$ would be relaxed when the BH mass is smaller or the spectral index is smaller.  This is because the radiation with higher $E_{\rm max}$ and/or flatter spectrum can heat up the ambient gas to higher temperature, which makes the conductive heating of the shielded region more efficient, and the heating timescale would be shorter when the BH mass is smaller since the size of the shielded region is smaller.  One can also see that when the ambient temperature is high enough the super-Eddington accretion would not be suppressed regardless of $E_{\rm max}$ because of the condition that the Compton heating timescale should be shorter than the accretion timescale at the Bondi radius (i.e., $R_{\rm Comp}^{\rm infall} < R_{\rm Bondi}$).  The critical ambient temperature would be higher when the central BH mass is larger.  From these results, one can say that a seed BH can grow more easily when the spectral index of the irradiation flux is softer and the mass of a seed BH is larger.

\section{Discussion}
\citet{2019MNRAS.488.2689T} investigated the effects of the irradiation spectra from the inner region of an accretion disk around a seed BH on its growth by performing two-dimensional multifrequency radiation hydrodynamical simulations, and they present the criterion required for super-Eddington accretion as $(M_{\rm BH}/10^5~M_{\odot})(n_{\infty}/10^4~{\rm cm}^{-3})\gtrsim 2.4(\langle \epsilon \rangle /100~{\rm eV})^{-5/9}$, where $\langle \epsilon \rangle$ is the mean energy of the irradiation flux.  In their work, anisotropic irradiation field and ionization heating are taken into account.  However, they did not take into account the heat transfer in non-radial directions.  Actually, when the ambient medium is fully ionized by the irradiation it is optically thin with respect to electron scattering and so we can neglect the non-radial heat transfer due to photon propagation.  In this study, we consider the heat conduction from an irradiated region to a shielded region (i.e., the equatorial region), which has not been considered in this context so far, and obtain the condition with which the medium surrounding an accreting seed BH would be heated up isotropically and the accretion rate would be suppressed.  We have shown that we cannot neglect the effects of heat conduction especially when the spectral index of the irradiation flux is as hard as $\alpha \simeq 1$ ($F_{\nu}\propto \nu^{-\alpha}$) and the maximum energy of irradiating photons is as high as $\gtrsim$ several tens of keV.  Fig. 6 depicts the dependence of $\langle \epsilon \rangle$ on $E_{\rm max}$ in cases with $\alpha=1.0$, $1.2$, and $1.4$.  One can see that even when the irradiation spectrum is as hard as $\alpha\gtrsim 1.1$ the dependence of $\langle \epsilon \rangle$ on $E_{\rm max}$ is very small, which means that not only $\langle \epsilon \rangle$ but also $E_{\rm max}$ is important for the criterion for super-Eddington accretion onto a BH.

Let us check if the heat conduction can affect the thermal property of an irradiated region significantly.  We can see that from Eq.(\ref{comptonheating}) the cooling rate per unit volume due to Compton scattering in the irradiated region is given by
\begin{eqnarray}
\Gamma_{\rm Comp}=\frac{L}{4\pi r^2}\frac{\sigma_T n_e}{m_e c^2} \cdot 4k_{\rm B}T_{\rm C},
\end{eqnarray}
while the cooling rate per unit volume due to heat conduction is given by
\begin{eqnarray}
\Gamma_{\rm cond}\simeq\frac{F_{\rm cond}}{r}\simeq \kappa_{\rm sp}\frac{T_{\rm C}}{r^2}.
\end{eqnarray}
Using the expression $L=\eta \dot{M}_{\rm Bondi}c^2 =6.71\times 10^{45}~{\rm erg}~{\rm s}^{-1}~\eta_{-2}T^{-3/2}_{\infty,4}n_{\infty,5}M_{{\rm BH},5}^2$, one can rewrite the condition $\Gamma_{\rm Comp} \gg \Gamma_{\rm cond}$ as
\begin{eqnarray}
M_{{\rm BH},5}\gg 3.6\times 10^{-8}T_{\rm C}^{5/2}\eta_{-2}^{-1}T_{\infty,4}^{3/2}n_{\infty,5}^{-2},
\end{eqnarray}
which is naturally fulfilled within our setup.  In other words, the conductive cooling is a subdominant process in the irradiated region compared to the Compton cooling.

How the spectral energy distribution depends on the accretion rate, black hole mass, and other environmental parameters is still under debate.  Conventionally, the X-ray emission from an accreting BH originates from a hot corona formed in the innermost region of an accretion flow, and the X-ray spectra of X-ray binaries and AGNs are well explained by some phenomenological model of coronae \citep{haardtmaraschi91, 1994ApJ...436..599S}.  The X-ray emission from a super-Eddington accreting BH is also supposed to come from a Comptonizing corona formed in the innermost region of an accretion disk.  However, one cannot predict its spectral index or maximum energy from the first principle because there is no established physical model of a disk corona.  In the context of ultraluminous X-ray sources, that are believed to harbor stellar-mass BHs accreting at super-Eddington rate, the coronal temperature inferred from observations is rather low ($\lesssim {\rm a~few}~10~{\rm keV}$), which implies that the coronae associated with super-Eddington accreting BHs generally have low temperature.  One can interpret this tendency by considering that the corona is formed by the radiation pressure-driven disk wind \citep{2021PASJ...73..630K}.  Whether one can apply this theoretical model to super-Eddington accreting BHs at high redshift is uncertain, and we leave it to future work.

The super-Eddington growth of seed BHs at high redshift will be testable using future observations.  For example, \cite{2022ApJ...931L..25I} show that super-Eddington accreting BHs in metal-poor galaxies at $z\gtrsim 8$ are detectable by James Webb Space Telescope (JWST), and that their colors affected by strong H$\alpha$ line emission can be the selection criteria to distinguish them with low-$z$ quasars.  \cite{2022arXiv220105300K} discuss how the intergalactic gas temperature would be modified by the radiation from super-Eddington accreting seed BHs at high redshifts, and which would affect the absorption feature of the cosmological 21 cm lines.  Future observations such as Square Kilometer Array (SKA) will give us more precise data of cosmological 21 cm lines, which will enable us to find some signatures of the super-Eddington growth of seed BHs.

Not only radiation or conduction, but also convection can transfer the heat in the irradiated region to the shielded region.  In fact, the density in the shielded region would be higher than that in the irradiated region because of the intense mass accretion from the Bondi radius, and there is velocity shear across the interface between these region.  Such a situation may make the boundary layer unstable with respect to Kelvin-Helmholtz instability, which would drive convective motion of the fluid in the boundary.  The quantitative discussion on the heat transfer due to this convection is beyond the scope of this work.

\begin{figure*}[tbp]
\begin{center}
\includegraphics[width=10cm]{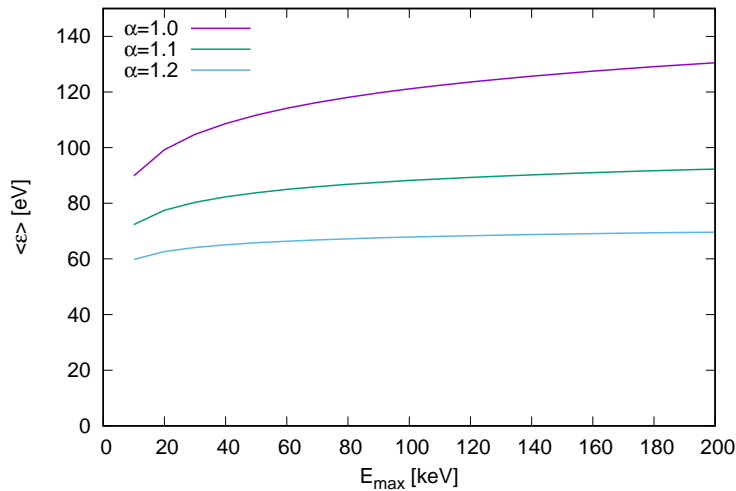}
\caption{Mean photon energy of the irradiation flux, $\langle \epsilon \rangle$, as a function of maximum photon energy, $E_{\rm max}$, in cases with $\alpha=1.0$, $1.1$, and $1.2$ are shown.  The minimum photon energy is fixed as $13.6~{\rm eV}$.}
\label{emean}
\end{center}
\end{figure*}

\section{Summary}

In this study we investigate how the conductive heat transfer affect on the growth of a seed BH at high redshift.  A seed BH can grow via Bondi accretion and the accretion flow likely forms a disk-like structure due to non-zero angular momentum of the ambient gas.  In such a case, as shown in some previous studies, the medium around the equatorial plane can be accreted at a super-Eddington rate because the X-ray irradiation from the vicinity of a BH is shielded by the geometrically-thick accretion flow.  However, if this medium is heated due to heat conduction from the adjacent hot region that is not shielded from the X-ray irradiation, the ambient gas becomes hot isotropically, which results in the suppression of Bondi accretion rate.  Such a non-radial heat transfer has not been investigated in previous studies.  In this study we investigate the necessary condition for this to occur and obtain a new criterion for super-Eddington accretion onto a seed BH.  We point out that the maximum photon energy and spectral index of the X-ray radiation from the vicinity of a BH is important for the criterion because it determines the Compton temperature of the surrounding medium, which is directly related to the heat conduction efficiency onto the shielded medium.  The spectral properties of X-ray radiation and their dependence of accretion rates are important in determining the effeiciency of the BH growth due to Bondi accretion, and so these should be explored both theoretically and observationally to investigating the SMBH formation in high redshift.

\section*{acknowledgement}
We are grateful to Shin Mineshige and Ramesh Narayan for their valuable comments.  This work is supported in part by the Hakubi project at Kyoto University, by JSPS KAKENHI Grant Number 22K03686 (N.K.), JP17H01131,  and by MEXT KAKENHI Grant Numbers JP20H04750, JP22H05270 (K.K.).


%
%\bibliography{sample631}{}
%\bibliographystyle{aasjournal}



\end{document}